\documentclass[nofootinbib,prd,aps,superscriptaddress,preprintnumbers,twocolumn,showpacs]{revtex4-1}
\usepackage[dvipdfmx]{graphicx}
\usepackage{amsmath,amssymb,amsthm,bm,bigdelim,multirow}
\usepackage{color}

\newcommand{\target}{{*}}

\newcommand{\kslash}{k\kern-1ex /}
\newcommand{\pslash}{p\kern-1ex /}
\newcommand{\qslash}{q\kern-1ex /}
\newcommand{\lslash}{l\kern-1ex /}
\newcommand{\sslash}{s\kern-1ex /}
\newcommand{\Dslash}{D\kern-1.2ex /}

\newcommand{\beqa}{\begin{eqnarray}}
\newcommand{\eeqa}{\end{eqnarray}}

\newcommand{\bd}{\begin{description}}
\newcommand{\ed}{\end{description}}
\newcommand{\la}{\langle}
\newcommand{\ra}{\rangle}
\newcommand{\ben}{\begin{eqnarray}}
\newcommand{\een}{\end{eqnarray}}
\newcommand{\nn}{\nonumber}
\def\lsim{\raise0.3ex\hbox{$<$\kern-0.75em\raise-1.1ex\hbox{$\sim$}}}
\def\gsim{\raise0.3ex\hbox{$>$\kern-0.75em\raise-1.1ex\hbox{$\sim$}}}
\def\simgt{\rlap{\lower 3.5 pt\hbox{$\mathchar \sim$}}\raise 2.0pt \hbox {$>$}}
\def\simlt{\rlap{\lower 3.5 pt\hbox{$\mathchar \sim$}}\raise 2.0pt \hbox {$<$}}

\newcommand{\crb}{\color{blue}}

\begin{document}
\title{Finite size effect on pseudoscalar meson sector in 2+1 flavor QCD\\ at the physical point}

\author{K.-I.~Ishikawa}
\affiliation{Graduate School of Science, Hiroshima University, Higashi-Hiroshima, Hiroshima 739-8526, Japan}

\author{N.~Ishizuka}
\affiliation{Center for Computational Sciences, University of Tsukuba, Tsukuba, Ibaraki 305-8577, Japan}

\author{Y.~Kuramashi}
\affiliation{Center for Computational Sciences, University of Tsukuba, Tsukuba, Ibaraki 305-8577, Japan}

\author{Y.~Nakamura}
\affiliation{RIKEN Center for Computational Science, Kobe, Hyogo 650-0047, Japan}

\author{Y.~Namekawa}
\affiliation{Faculty of Pure and Applied Sciences, University of Tsukuba, Tsukuba, Ibaraki 305-8571, Japan}

\author{Y.~Taniguchi}
\affiliation{Center for Computational Sciences, University of Tsukuba, Tsukuba, Ibaraki 305-8577, Japan}

\author{N.~Ukita}
\affiliation{Center for Computational Sciences, University of Tsukuba, Tsukuba, Ibaraki 305-8577, Japan}

\author{T.~Yamazaki}
\affiliation{Faculty of Pure and Applied Sciences, University of Tsukuba, Tsukuba, Ibaraki 305-8571, Japan}

\author{T.~Yoshi\'e}
\affiliation{Center for Computational Sciences, University of Tsukuba, Tsukuba, Ibaraki 305-8577, Japan}


\collaboration{PACS Collaboration}

\begin{abstract}
We investigate the finite size effect on pseudoscalar meson masses and decay constants using a subset of  the ``PACS10'' configurations which are generated keeping the space-time volumes over (10 fm$)^4$ in 2+1 flavor QCD at the physical point. We have tried two kinds of analyses, fixing $\kappa$ values or measured axial Ward identity quark masses. Comparing the results on (5.4 fm$)^4$ and (10.8 fm$)^4$ lattices, we have found a sizable finite size effect on the pseudoscalar meson sector in the former analysis: a 2.1{(8)}\%, 4.8{(1.6)}\%, and 0.36{(31)}\% finite size effect on $m_\pi$, $m_{\rm ud}$, and $f_\pi$, respectively, on the (5.4 fm$)^4$ lattice. For the latter analysis, the finite size effect on the pseudoscalar meson decay constants is 0.66{(33)}\% for $f_\pi$, 0.26{(13)}\% for $f_K$,  {and 0.40(32)\% for $f_K/f_\pi$. These values with two-sigma error bars are consistent with the predictions from the full one-loop SU(3) chiral perturbation theory, which are 0.20\% for $f_\pi$, 0.08\% for $f_K$, and 0.13\% for $f_K/f_\pi$.} The finite size effect on the pseudoscalar meson masses is hardly detected under the current statistical precision.
\end{abstract}
\date{\today}

\preprint{UTHEP-722, UTCCS-P-114, HUPD-1805}

\maketitle

\newpage
\section{Introduction}

Lattice QCD simulations on very large lattices, which are named master-field simulations by L{\"u}scher~\cite{master-field}, have various potential advantages: the statistical errors decrease thanks to the geometrical symmetries of the lattice~\cite{ukita18}, the accessible minimum momentum is reduced in proportion to $1/L$ with $L$ the lattice extent, and we could be free from the finite size effect on the low energy properties of the baryons which is potentially severe discussed in Ref.~\cite{nff_kconf}. Since the systematic study of the finite size effect demands huge computational cost, lattice QCD practitioners have been heavily depending on the analytic estimations, so far~\cite{fse_analytic,fse_chpt}.

The PACS Collaboration is now generating 2+1 flavor QCD configurations on very large lattices over (10 fm$)^4$ at the physical point using the Wilson-type quarks. These are called ``PACS10'' configurations. This project is the successor to the PACS-CS project which mainly focused on reducing the up-down quark masses up to the physical point~\cite{pacs-cs1,pacs-cs2}. Since the PACS10 configurations have very large physical volumes, they should provide us a good opportunity to investigate the finite size effect in 2+1 flavor QCD.  We have made a finite size study employing (10.8 fm$)^4$ and (5.4 fm$)^4$ lattices at a cutoff of $a^{-1}\approx 2.3\ $GeV; the latter of which is a typical lattice size in current 2+1 flavor lattice QCD simulations at the physical point. The same hopping parameter $\kappa$ is chosen for both lattices. We have performed two types of analyses: one is a comparison between the results for the pseudoscalar (PS) meson sector at the same hopping parameter on both lattices. The other is a comparison at the same axial Ward identity (AWI) quark masses on both lattices, where the quark masses on the smaller lattice are adjusted to those on the larger one by the reweighting method~\cite{pacs-cs2}. We have observed different types of finite size effects in two analyses. It is reasonable to make a comparison with the chiral perturbation theory (ChPT) in the second analysis with the quark mass fixed on both lattices. 

In this paper, we present details of two different types of analyses for the finite size effect on the pseudoscalar meson sector. {The first analysis with the fixed $\kappa$ shows about 5\% finite size effect for the axial Ward identity (AWI) quark mass $m_{\rm ud}$. The results in the second analysis with the fixed AWI quark mass are compared with the ChPT predictions. We find that the finite size effect on the PS meson decay constants is at most 0.7\%}. As for the PS meson masses, it is difficult to detect the finite size effect beyond the statistical errors. 

This paper is organized as follows.
The simulation details are given in Sec.~\ref{sec:simulation}.
In Sec.~\ref{sec:results}, we present the results obtained with two kinds of analyses for the finite size effect and compare the results with the ChPT predictions. Our conclusions and outlook are summarized in Sec.~\ref{sec:conclusion}.

\section{Simulation details}
\label{sec:simulation}

\subsection{Configuration generation}
\label{subsec:generation}

Following Ref.~\cite{k-config}, we have generated 2+1 flavor QCD gauge configurations employing the stout-smeared $O(a)$-improved Wilson-clover quark action and Iwasaki gauge action~\cite{iwasaki} on $V=L^3\times T=128^4$ and $64^4$ lattices at $\beta=1.82$.
The corresponding lattice spacing is $a=0.08457(67)$ fm [$a^{-1}=2.333(18)$] GeV~\cite{k-config} so that the physical lattice volumes reach (10.8 fm$)^4$ and (5.4 fm$)^4$, respectively. We use the stout smearing parameter $\rho = 0.1$~\cite{Morningstar:2003gk}, and the number of the smearing is six.
We adopt a value of the improvement coefficient $c_{\rm SW}=1.11$, which is nonperturbatively determined by the Schr{\"o}dinger functional (SF) scheme~\cite{csw_sf}. The hopping parameters for the light (up-down) and strange quarks $(\kappa_{\rm ud},\kappa_{\rm s})=(0.126117,0.124902)$ are carefully adjusted to yield the physical pion and kaon masses $(m_\pi,m_K)$=(135.0MeV, 497.6MeV) on a $128^4$ lattice within the cutoff error.  

The degenerated up-down quarks are simulated with the domain-decomposed HMC (DDHMC) algorithm~\cite{ddhmc} both on $64^4$ and $128^4$ lattices. The ud quark determinant is separated into the UV and IR parts after the even-odd preconditioning. We also apply the twofold mass preconditioning~\cite{massprec1, massprec2} to the IR part by splitting it into $\tilde F_{\rm IR}$, $F_{\rm IR}^{\prime}$, and $F_{\rm IR}^{\prime\prime}$. This decomposition is controlled by two additional hopping parameters:
$\kappa^\prime_{\rm ud}=\rho_1\kappa_{\rm ud}$ with $\rho_1=0.9997$ and 
$\kappa^{\prime\prime}_{\rm ud}=\rho_1\rho_2\kappa_{\rm ud}$ with $\rho_2=0.9940$. ${\tilde F}_{\rm IR}$ is derived from the action preconditioned with
$\kappa^\prime_{\rm ud}$. The ratio of two preconditioners with 
$\kappa^\prime_{\rm ud}$ and $\kappa^{\prime\prime}_{\rm ud}$ gives
$F^{\prime}_{\rm IR}$.  
$F^{\prime\prime}_{\rm IR}$ is from the heaviest preconditioner 
with $\kappa^{\prime\prime}_{\rm ud}$. 
In the end, the force terms  consist of  the gauge force $F_{\rm g}$, the UV force $F_{\rm UV}$, and the three IR forces $F_{\rm IR}^{\prime\prime}$, $F_{\rm IR}^{\prime}$, and $\tilde F_{\rm IR}$. 
We adopt the multiple time scale integration scheme~\cite{sexton}  in the molecular dynamics (MD) steps. 
The associated step sizes are controlled by a set of integers $(N_0,N_1,N_2,N_3,N_4)$: $\delta\tau_{\rm g}=\tau/N_0 N_1 N_2 N_3 N_4,$ 
$\delta\tau_{\rm UV}=\tau/N_1 N_2 N_3 N_4,$ 
$\delta\tau^{\prime\prime}_{\rm IR}=\tau/N_2 N_3 N_4,$ 
$\delta\tau^{\prime}_{\rm IR}=\tau/N_3 N_4,$ 
$\delta{\tilde \tau}_{\rm IR}=\tau/N_4$ with $\tau=1.0$.
Our choice of $(N_0,N_1,N_2,N_3,N_4)=(8,2,2,2,12)$ and $(8,2,2,2,22)$ for the $64^4$ and $128^4$ lattices results in 80\%  and 72\% acceptance rates, respectively.

The strange quark on a $64^4$ lattice is simulated with the UVPHMC algorithm~\cite{fastMC,Frezzotti:1997ym,phmc,ishikawa_lat06}, where the action is UV filtered~\cite{Alexandrou:1999ii} after the even-odd preconditioning without domain decomposition. We set the step size as
$\delta\tau_{\rm s}=\delta\tau^{\prime\prime}_{\rm IR}$
according to our observation  $||F_{\rm s}||\approx ||F_{\rm IR}^{\prime\prime}||$. 
This algorithm is made exact
by correcting the polynomial approximation with the 
global Metropolis test~\cite{NoisyMetropolisMB} at the end of each trajectory.
We find that the choice of $N_{\rm poly}=350$ 
yields 99\% acceptance rate. 
On the other hand, we employ the RHMC algorithm~\cite{rhmc} to save the memory in simulating the strange quark on a $128^4$ lattice, choosing the force approximation range of [min,max]=[0.00025,1.85] with $N_{\rm RHMC}=8$ and $\delta\tau_{\rm s}=\delta\tau^{\prime\prime}_{\rm IR}$ for the step size. 
After thermalization, we generate 2000 and 200 trajectories on $64^4$ and $128^4$ lattices, respectively, and calculate hadronic observables at every ten trajectories. 

\begin{figure}[h]
\begin{center}
\includegraphics[width=75mm,angle=0]{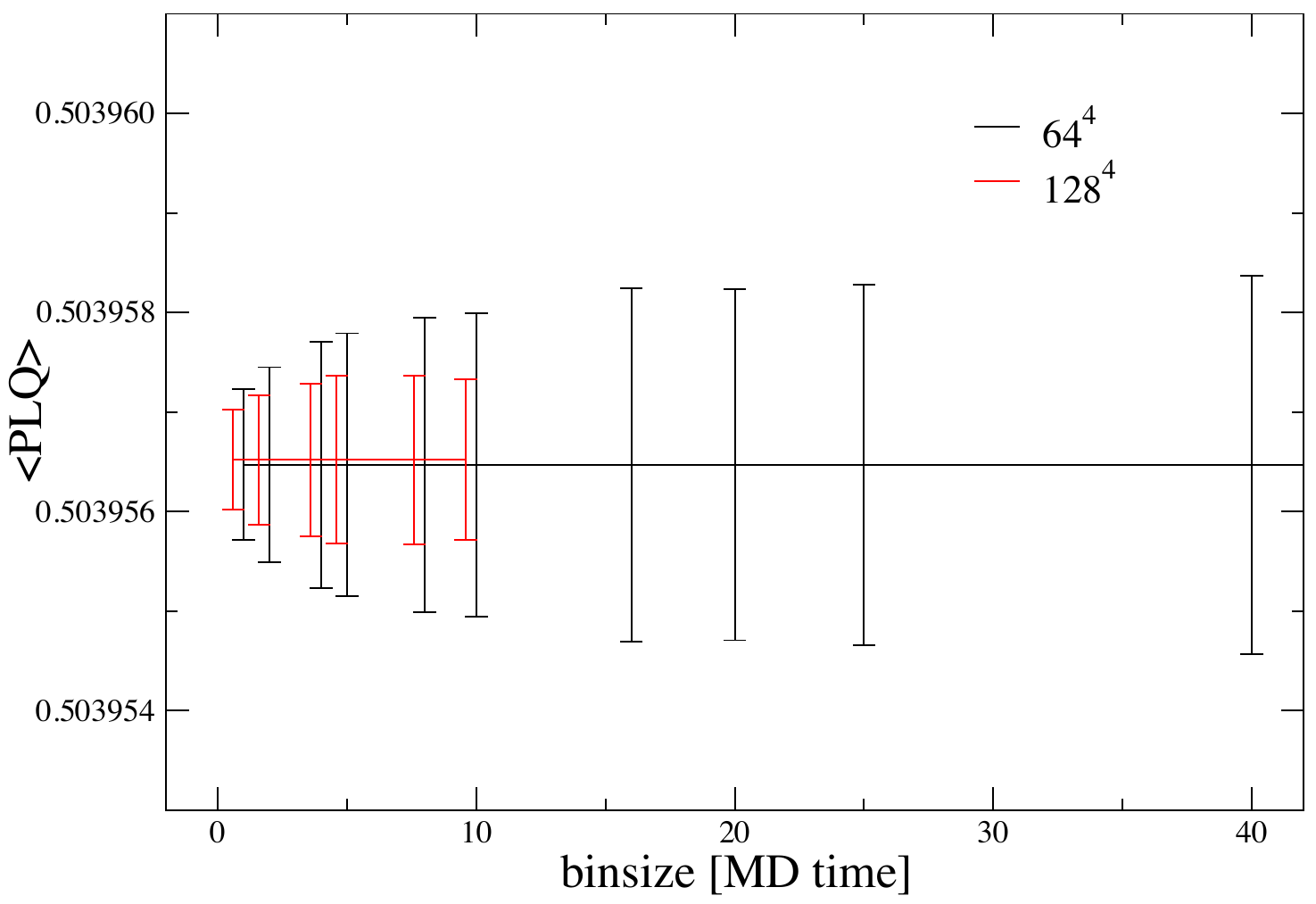}\\
\end{center}
\vspace{-5mm}
\caption{Binsize dependence in jackknife analysis for the plaquette values on $64^4$ and $128^4$ lattices. Data for $128^4$ lattice are horizontally shifted for clarification.}
\label{fig:plaq}
\end{figure}

In Fig.~\ref{fig:plaq}, we present a jackknife analysis for the plaquette value averaged over the space-time volume on  $64^4$ and $128^4$ lattices. The central values of the plaquette on both lattices show good consistency. An important observation is that the magnitude of the error on $128^4$ lattice is saturated around 5 MD time, while the error for  $64^4$ lattice needs about 20 MD time to be independent of the binsize. This is because the correlation between successive configurations is reduced in proportion to $1/\sqrt{V}$ thanks to the stochastic locality~\cite{master-field}, once the physical extent of the lattice goes beyond the relevant scale for the target physical observable. We also point out that the error on $128^4$ lattice is much smaller than that on $64^4$ lattice.

\subsection{Reweighting technique}
\label{subsec:reweighting}

When we investigate the finite size effect equalizing the AWI quark mass between $64^4$ and $128^4$ lattices, we need to make a tiny shift of the hopping parameters on the smaller lattice employing the reweighting technique.
The reweighting factor for the ud quark is evaluated with a stochastic method introducing 
 a set of independent Gaussian random noises $\eta_i$ $(i=1,\dots,N_\eta)$,
\begin{eqnarray}
 {\rm det}[W_{\rm ud}^2] &=&
 \left[\lim_{N_\eta\rightarrow \infty}
 \frac{1}{N_\eta} \sum_{i=1}^{N_\eta} 
  {\rm e}^{-\vert W_{\rm ud}^{-1}\eta_i\vert^2 +\vert \eta_i\vert^2}\right],
 \label{eq:rwfactor_ud}\\
 W_{\rm ud} &=& \frac{D(\kappa_{\rm ud}^*)}{D(\kappa_{\rm ud})},
\label{eq:w_ud}
\end{eqnarray}
where $D(\kappa_{\rm ud}^*)$ is the Wilson-Dirac matrix with a target hopping parameter $\kappa_{\rm ud}^*$. For the strange quark, we employ the square root trick, 
\begin{eqnarray}
 {\rm det}[W_{\rm s}] &=&
 \left[\lim_{N_\eta\rightarrow \infty}
 \frac{1}{N_\eta} \sum_{i=1}^{N_\eta} 
  {\rm e}^{-\vert W_{\rm s}^{-1}\eta_i\vert^2 +\vert \eta_i\vert^2}\right]^{\frac{1}{2}},
 \label{eq:rwfactor_s}\\
 W_{\rm s} &=& \frac{D(\kappa_s^*)}{D(\kappa_s)}.
\label{eq:w_s}
\end{eqnarray}
To reduce the fluctuation in the stochastic evaluation (\ref{eq:rwfactor_ud}) and (\ref{eq:rwfactor_s}), we employ the determinant breakup technique~\cite{pacs-cs2, hasenfratz, RBC_UKQCD}, in which 
the interval between $\kappa_{q}$ and $\kappa_{q}^*$ is divided into $N_B^q$
 subintervals for $q={\rm ud}, {\rm s}$.
 
\subsection{Measurement of hadronic observables}
\label{subsec:measurement}

Since we are interested in the PS meson state  with zero spatial momentum projection, we use the following local PS operator and axial vector current:
\ben
P(t)&=&\sum_{\vec x}{\bar q}_f({\vec x},t)\gamma_5 q_g({\vec x},t), \\
A_{\mu}(t)&=&\sum_{\vec x}{\bar q}_f({\vec x},t)\gamma_\mu\gamma_5 q_g({\vec x},t), 
\een 
where $f$ and $g$ ($f,g={\rm ud},{s}$) label the valence quark flavors. We do not take account of the $O(a)$ improvement of the axial vector current, because the coefficient $c_A$ was evaluated to be consistent with zero within the statistical error~\cite{csw_sf}{\crb \footnote{The $O(a)$ effect of the $c_A$ term should be canceled in comparison of the axial vector currents on different volumes.}}. 
The correlation functions are calculated by employing the wall source method without gauge fixing~\cite{ws_ngf}. We make eight measurements in each space-time direction, which amount to $8\times 4=32$ measurements in total for each gauge configuration both on  $64^4$ and $128^4$ lattices. 
The statistical errors are estimated with the jackknife method. After investigating the bin size dependence we have chosen 50 and 10 trajectories for $64^4$ and $128^4$ lattices, respectively. The correlation between successive configurations on $128^4$ lattice is reduced by about $\sqrt{128^4/64^4}=4$ compared to that on the $64^4$ lattice.

The PS meson masses are extracted from the correlation function of $\la P(t)P^\dagger(0)\ra$, whose $t$ dependence is given by
\ben
\la P(t)P^\dagger(0)\ra&=&L^3 C_{PP}W(m_{\rm PS}T) \label{eq:pp}\\
&&\times \left[\exp(-m_{\rm PS}t)+\exp(-m_{\rm PS}(T-t))\right], \nn
\een
where $W(m_{\rm PS}T)$ with $T=64$ and 128 denotes the contribution from the PS meson propagation wrapping around the lattice in the time direction,
\ben
W(m_{\rm PS}T)&=&1+\exp(-m_{\rm PS}T)+\exp(-2m_{\rm PS}T)+\cdots \nn\\
&=&\frac{1}{1-\exp(-m_{\rm PS}T)}.
\een

The PS meson decay constant is defined by
\ben
Z_A\sqrt{2\kappa_f}\sqrt{2\kappa_g}
\left\vert\la 0|A_4|{\rm PS}\ra\right\vert 
=f_{\rm PS}m_{\rm PS}
\een
with $|{\rm PS}\ra$ the PS meson state at rest and $Z_A$ the renormalization factor of the axial current. We evaluate $f_{\rm PS}$ with the following combination:
\ben
f_{\rm PS}=Z_A\sqrt{2\kappa_f}\sqrt{2\kappa_g}
\frac{\sqrt{2}\left\vert C_{A P}\right\vert}
{\sqrt{ m_{\rm PS}  \left\vert C_{PP}\right\vert}},
\een
where we extract $m_{\rm PS}$, $C_{A P}$, $C_{PP}$ from a simultaneous fit of Eq.~(\ref{eq:pp}) and 
\ben
 \la A_4(t)P^\dagger(0)\ra&=&L^3 C_{AP}W(m_{\rm PS}T) \\
&&\times \left[\exp(-m_{\rm PS}t)-\exp(-m_{\rm PS}(T-t))\right]\nn
\label{eq:ap}
\een
with a common fit range of $[t_{\rm min},t_{\rm max}]=[17,60]$ and $[20,60]$ for $\pi$ and $K$ mesons, respectively, on the $128^4$ lattice and $[t_{\rm min},t_{\rm max}]=[17,30]$ and $[20,30]$ on the $64^4$ lattice. Since we are interested in the finite size effect, we use the central value of $Z_A=0.9650(68)(95)$, which was nonperturbatively determined by the Schr{\"o}dinger functional scheme~\cite{za_sf}.

We define the bare AWI quark mass as
\ben
m^{\rm AWI}_f+ m^{\rm AWI}_g=\frac{\langle 0 |\nabla_4
  A_4|{\rm PS} \rangle}{\langle 0| P | {\rm PS}\rangle}.
\label{mAWTI}
\een
The AWI quark masses are evaluated with
\ben
m^{\rm AWI}_f+m^{\rm AWI}_g&=&m_{\rm PS}
\left\vert \frac{C_{AP}}{C_{PP}}\right\vert,
\een
where $m_{\rm PS}$, $C_{A P}$, $C_{PP}$ are extracted from a simultaneous fit of Eqs.~(\ref{eq:pp}) and  (\ref{eq:ap}) with $[t_{\rm min},t_{\rm max}]=[17,60]$ and $[20,60]$ for $\pi$ and $K$ mesons, respectively, on the $128^4$ lattice and $[t_{\rm min},t_{\rm max}]=[17,30]$ and $[20,30]$ on the $64^4$ lattice.

\section{Numerical results}
\label{sec:results}

\subsection{Finite size effect at the fixed hopping parameters}
\label{subsec:fse_kappa}

\begin{figure}[h]
\begin{center}
\includegraphics[width=75mm,angle=0]{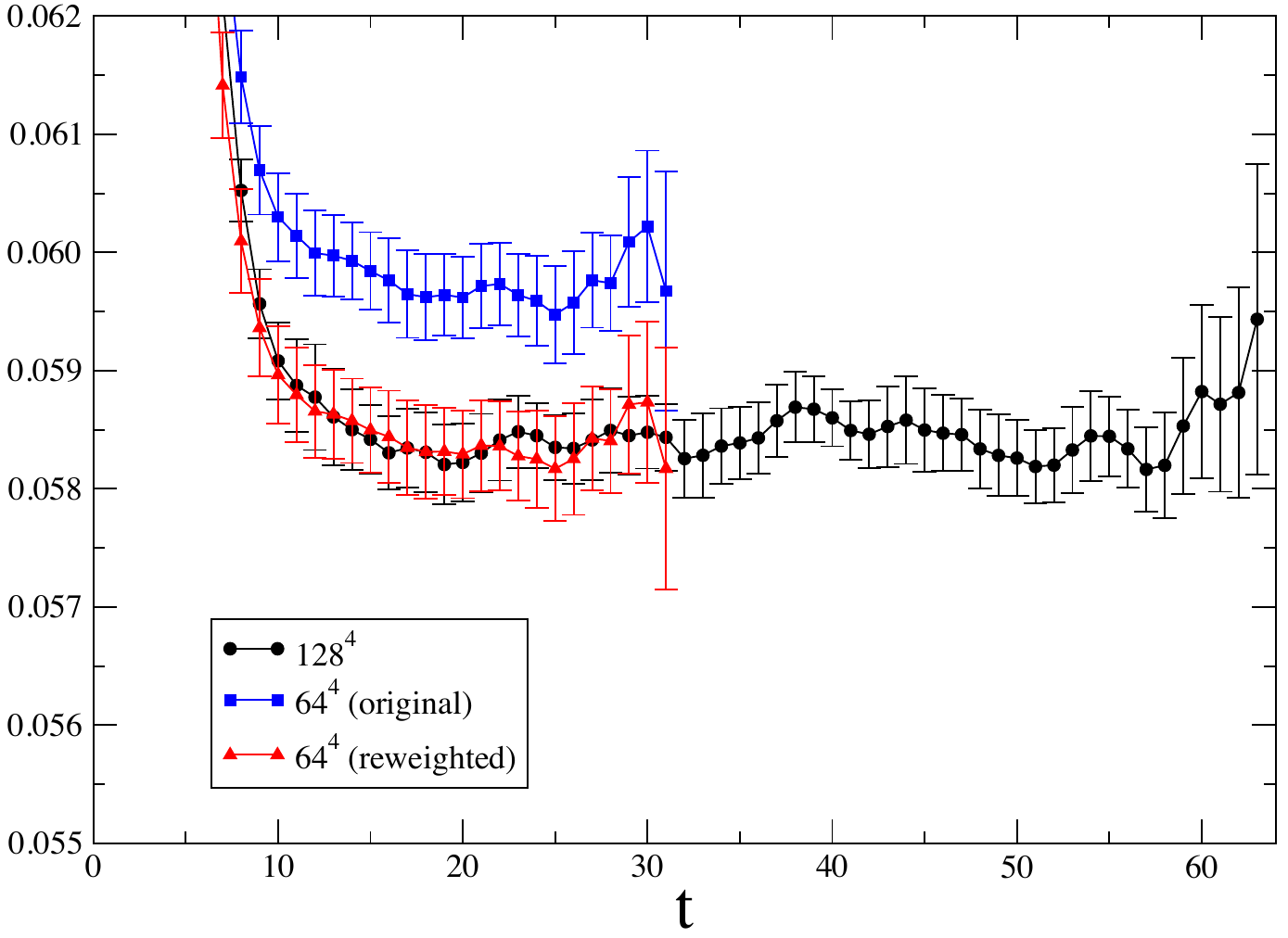}\\
\includegraphics[width=75mm,angle=0]{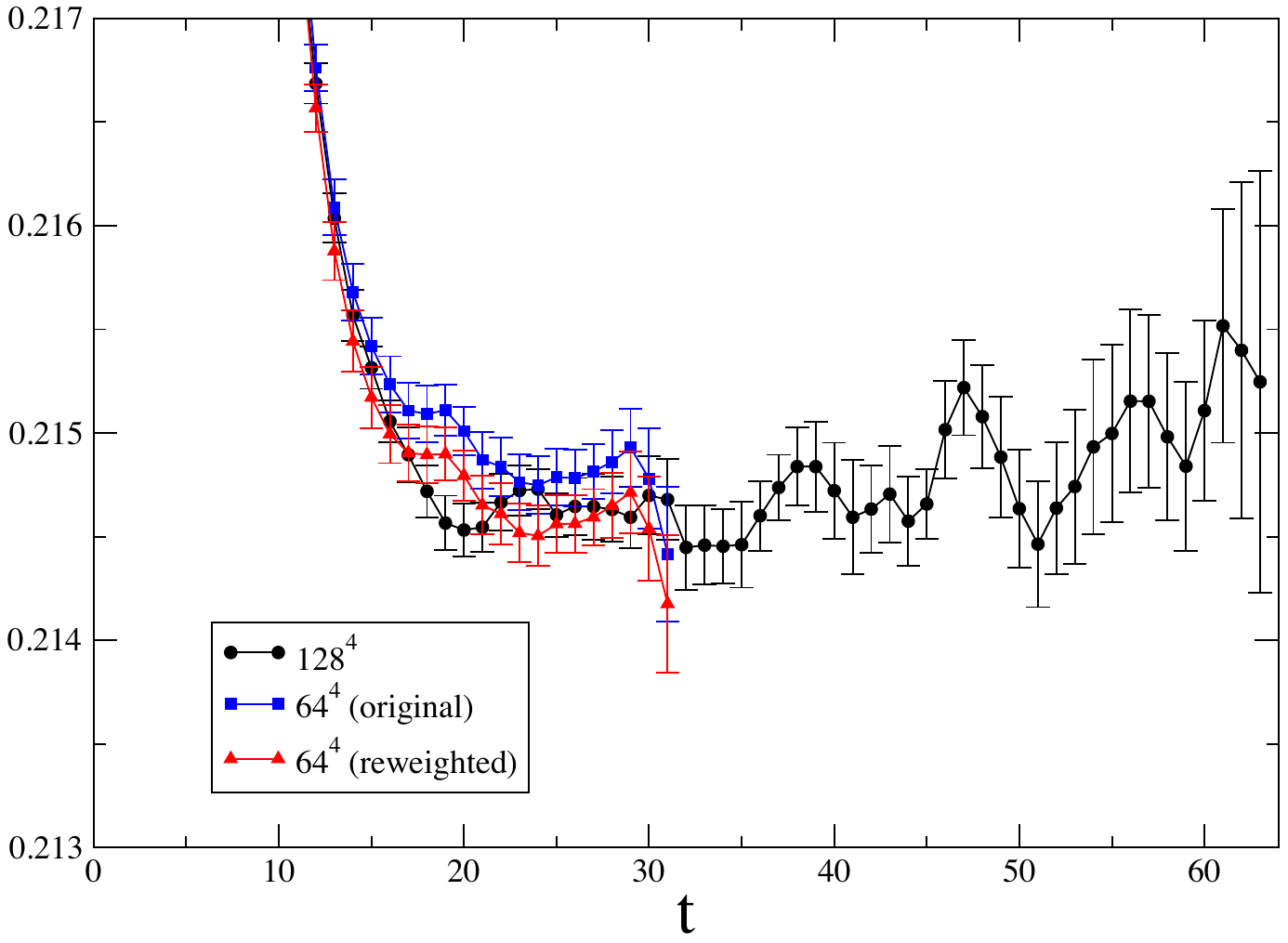}
\end{center}
\vspace{-5mm}
\caption{Comparison of the effective masses for $\pi$ (top) and $K$ (bottom) mesons on $64^4$ and $128^4$ lattices.}
\label{fig:effmass}
\end{figure}

\begin{figure}[h]
\begin{center}
\includegraphics[width=75mm,angle=0]{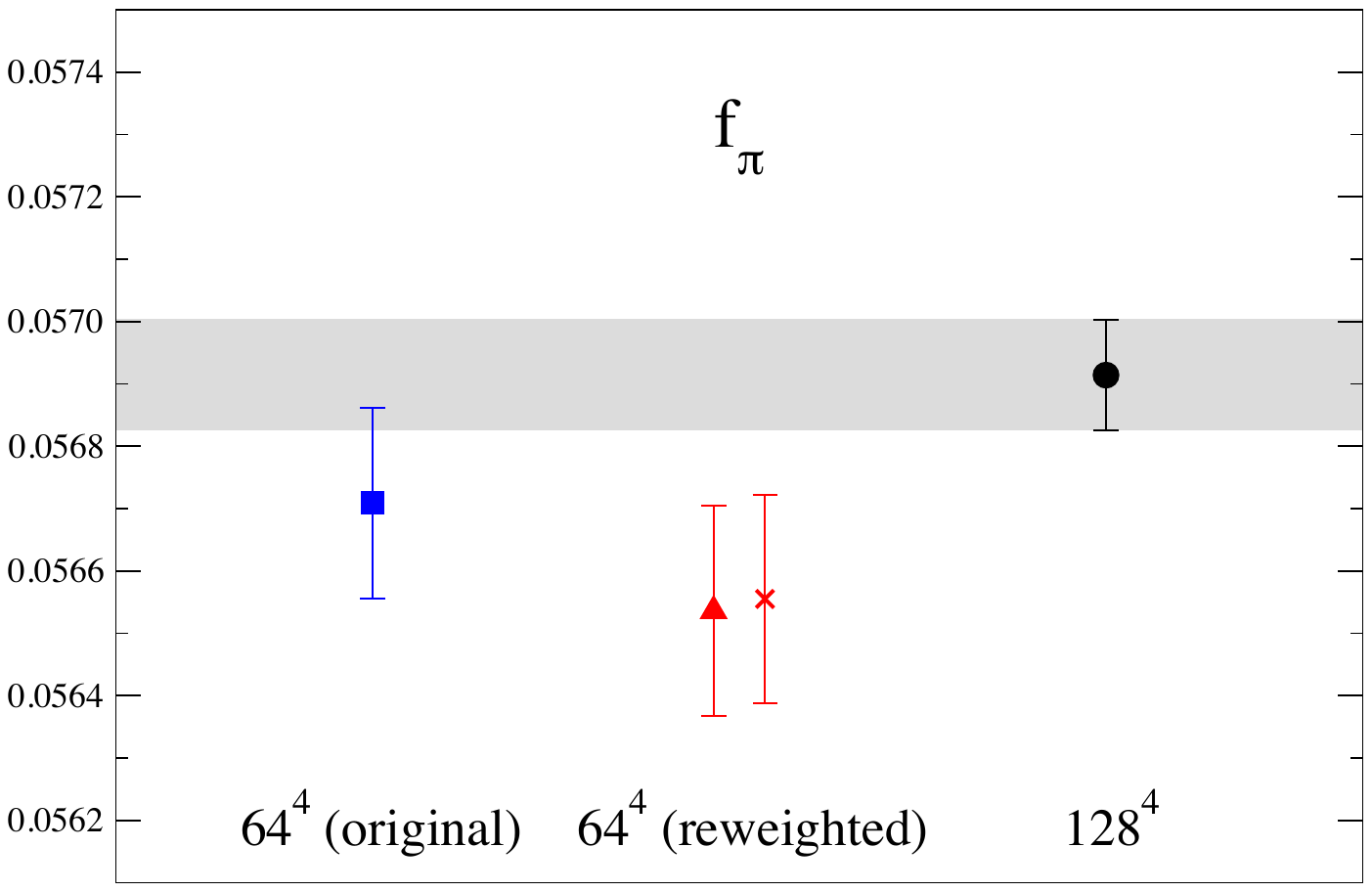}\\
\includegraphics[width=75mm,angle=0]{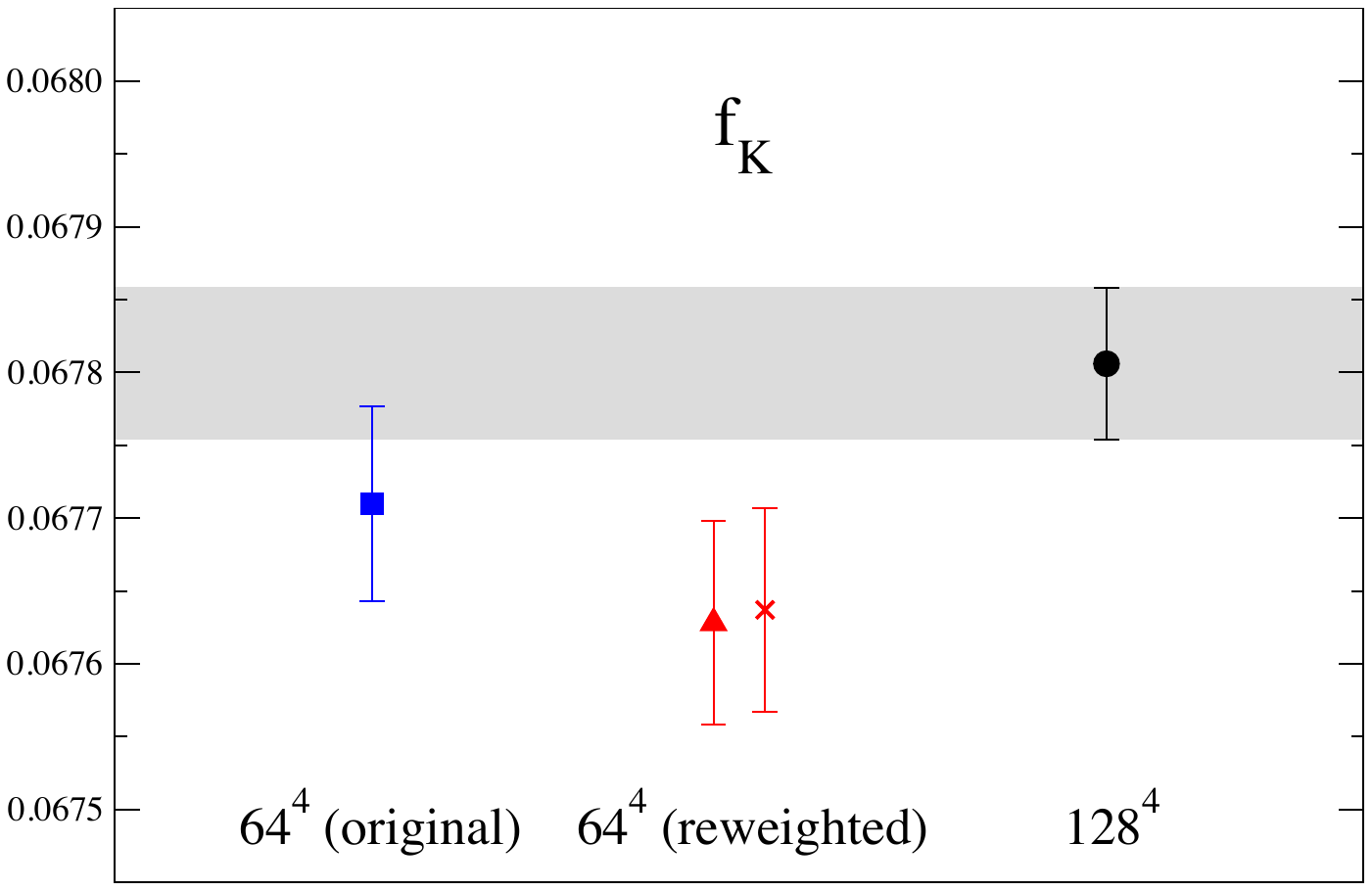}\\
\includegraphics[width=75mm,angle=0]{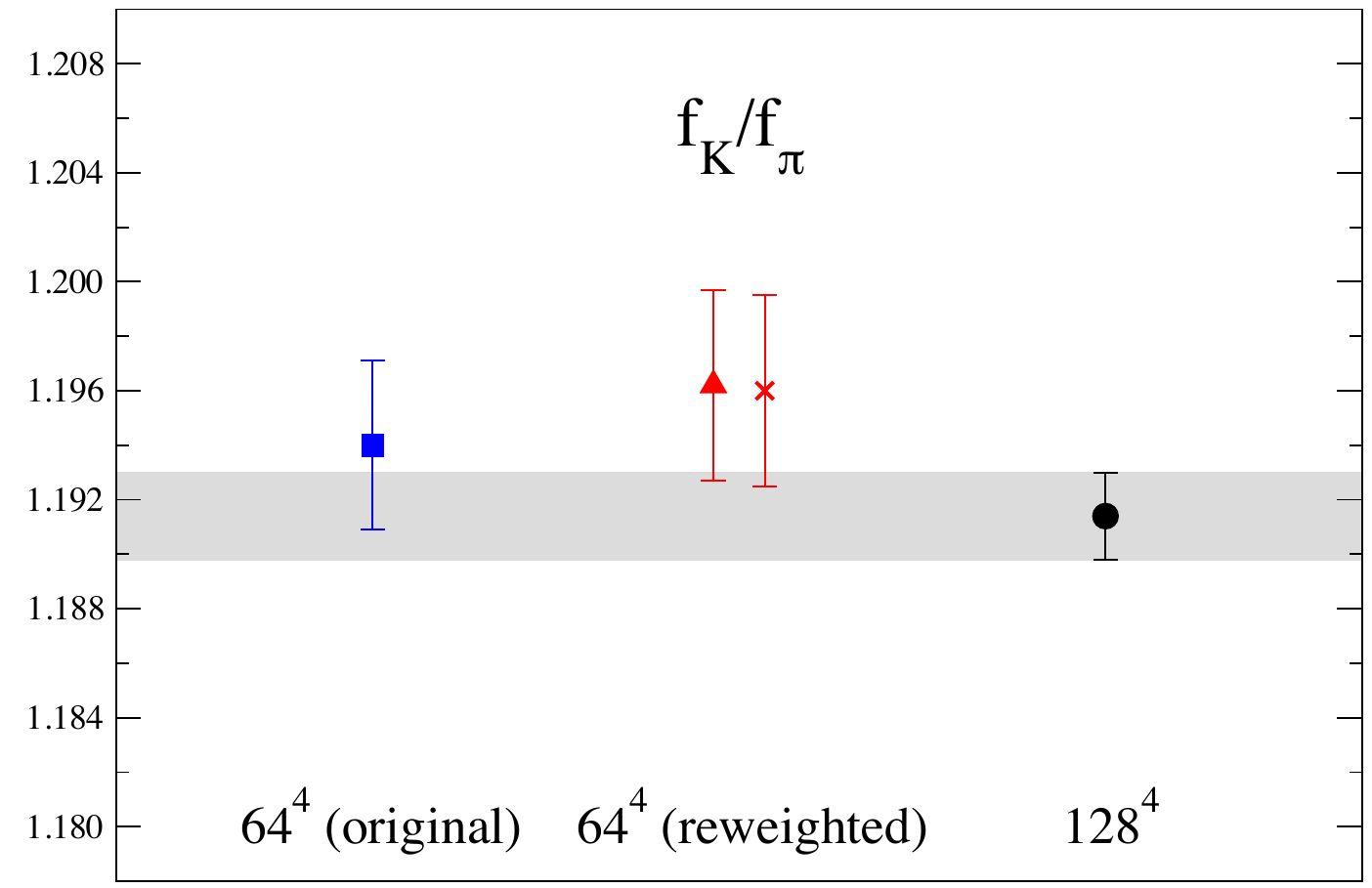}
\end{center}
\caption{Comparison of the decay constants for $\pi$ (top), $K$ (middle) mesons, and $f_K/f_{\pi}$ (bottom) on $64^4$ and $128^4$ lattices. A shaded band denotes the error band of the $128^4$ result. {A red triangle symbol denotes the result reweighted to the point where the AWI quark masses on $64^4$ lattice are equal to those on $128^4$ lattice, while a red cross symbol represents the interpolated result onto the point where the reweighted pion mass reproduces the ChPT prediction (see Sec.~\ref{subsec:chpt}).}}
\label{fig:decayconst}
\end{figure}

We first compare the results on $64^4$ and $128^4$ lattices at the same hopping parameters $(\kappa_{\rm ud},\kappa_{\rm s})=(0.126117,0.124902)$. Figure~\ref{fig:effmass} shows the effective masses for the PS mesons. We observe that the effective $\pi$ meson mass on the $64^4$ lattice is clearly heavier than that on the $128^4$ lattice beyond the error bars. On the other hand, the effective mass for the $K$ meson shows little finite size effect. In Table~\ref{tab:fitmass}, we summarize the fit results for the PS meson masses $m_{\pi,K}$, choosing the fit range of $[t_{\rm min},t_{\rm max}]=[17,60]$ and $[20,60]$ for $\pi$ and $K$ mesons, respectively, on the $128^4$ lattice and $[t_{\rm min},t_{\rm max}]=[17,30]$ and $[20,30]$ on the $64^4$ lattice. The deviation in the $\pi$ meson channel is found to be {2.1(8)\%}. We also list the AWI quark masses $m_{\rm ud,s}$ in Table~\ref{tab:qmass}. The ud quark mass on the $64^4$ lattice is heavier than that on the $128^4$ lattice by 4.8{(1.6)}\%, in accordance with the finite size effect found for $m_\pi$. Essentially, what makes the $\pi$ meson mass heavier on $64^4$ lattice is the increment of the ud quark mass, which is caused by the shift of the critical kappa $\kappa_{\rm c}$ due to the finite size effect.
For the decay constants, we plot the results on $64^4$ and $128^4$ lattices in Fig.~\ref{fig:decayconst}, which are obtained by the method explained in Sec.~\ref{subsec:measurement}. Their numerical values are presented in Table~\ref{tab:fitdecayconst}. The small [0.36{(31)}\%] finite size effect is observed in the $\pi$ meson channel, though it is hardly detected in the $K$ meson channel. It should be noted that our results show an expected feature from ChPT that the finite size effect makes the values of the decay constants smaller as the spatial volume decreases. 

\begin{table}[t!]
\caption{Fit results for the PS meson masses.}
\begin{center}
\begin{tabular}{ccccccc}
\hline
Lattice size & $m_{\pi}$ & $m_{K}$ \\\hline
 $128^4$ & 0.058431(275) & 0.214677(083)  \\
 $64^4$ (original)  & 0.059647(349) & 0.214813(110)  \\
 $64^4$ (reweighted)  & 0.058311(376) & 0.214586(114) \\\hline
\end{tabular}
\end{center}
\label{tab:fitmass}
\end{table}%

\begin{table}[t!]
\caption{Results for the AWI quark masses.}
\begin{center}
\begin{tabular}{ccccccc}
\hline
Lattice size & $m_{\rm ud}$ & $m_{\rm s}$ & $m_{\rm s}/m_{\rm ud}$\\\hline
 $128^4$ & 0.001366(14) & 0.037983(06) & 27.80(29) \\
 $64^4$ (original)  & 0.001432(17) & 0.038006(09) & 26.54(31) \\
 $64^4$ (reweighted)  & 0.001367(18) & 0.037998(10) & 27.79(37) \\\hline
\end{tabular}
\end{center}
\label{tab:qmass}
\end{table}%

\begin{table}[t!]
\caption{Results for the PS meson decay constants.}
\begin{center}
\begin{tabular}{ccccccc}
\hline
Lattice size & $f_{\pi}$ & $f_{K}$ & $f_{K}/f_{\pi}$\\\hline
 $128^4$ & 0.056914(088) & 0.067806(52) & 1.1914(16) \\
 $64^4$ (original)  & 0.056709(153) & 0.067710(67) & 1.1940(31) \\
 $64^4$ (reweighted)  & 0.056536(169) & 0.067628(70) & 1.1962(35) \\\hline
\end{tabular}
\end{center}
\label{tab:fitdecayconst}
\end{table}%

\subsection{Finite size effect at the fixed AWI quark masses}
\label{subsec:fse_qmass}

\begin{figure}[h]
\begin{center}
\includegraphics[width=75mm,angle=0]{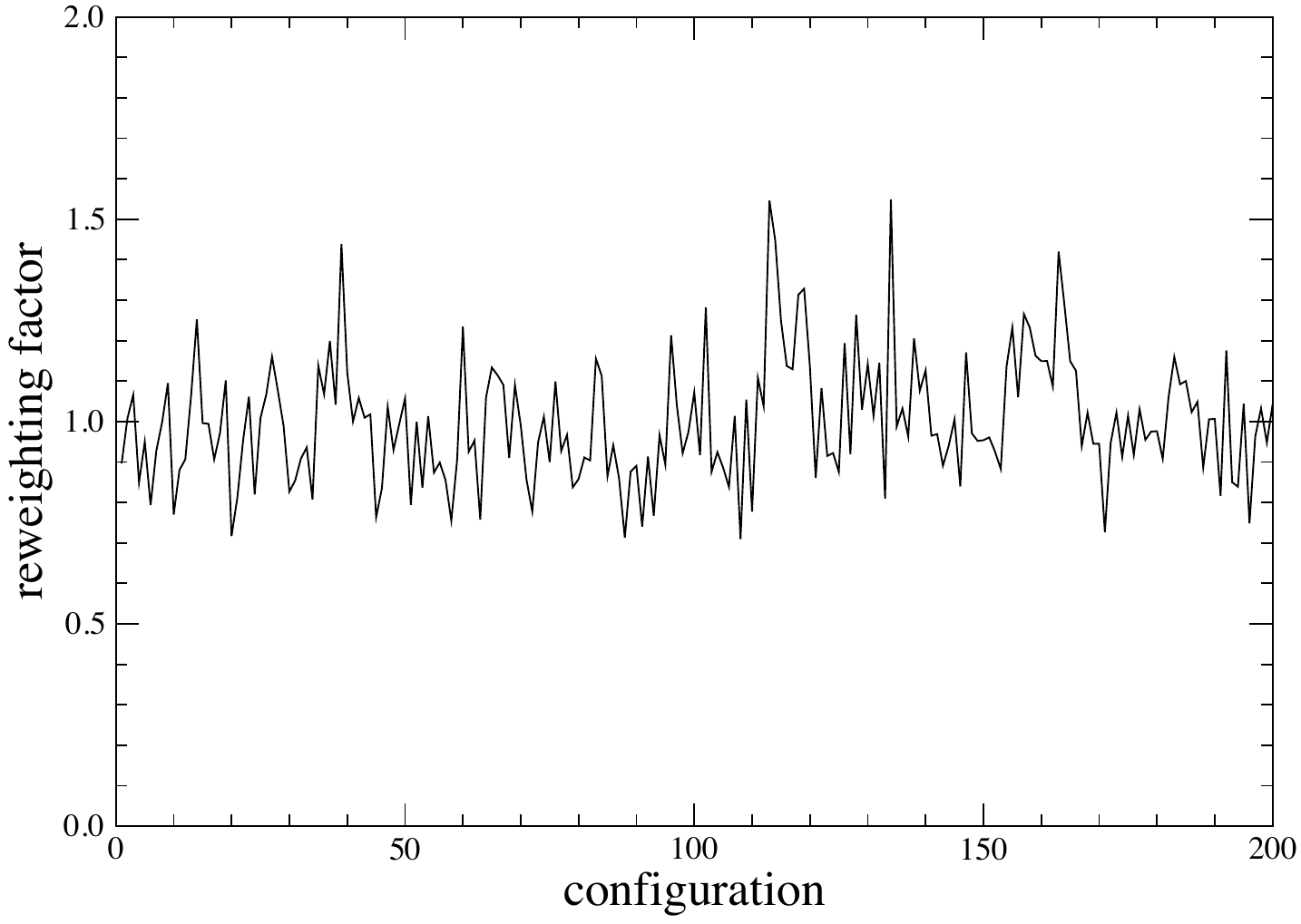}
\end{center}
\caption{Configuration dependence of the reweighting factor from 
$(\kappa_{\rm ud},\kappa_{\rm s})=(0.126117,0.124902)$ to 
$(\kappa_{\rm ud}^\target,\kappa_{\rm s}^\target)=(0.126119,0.124902)$.}
\label{fig:rwfactor_c}
\end{figure}

\begin{figure}[h]
\begin{center}
\includegraphics[width=75mm,angle=0]{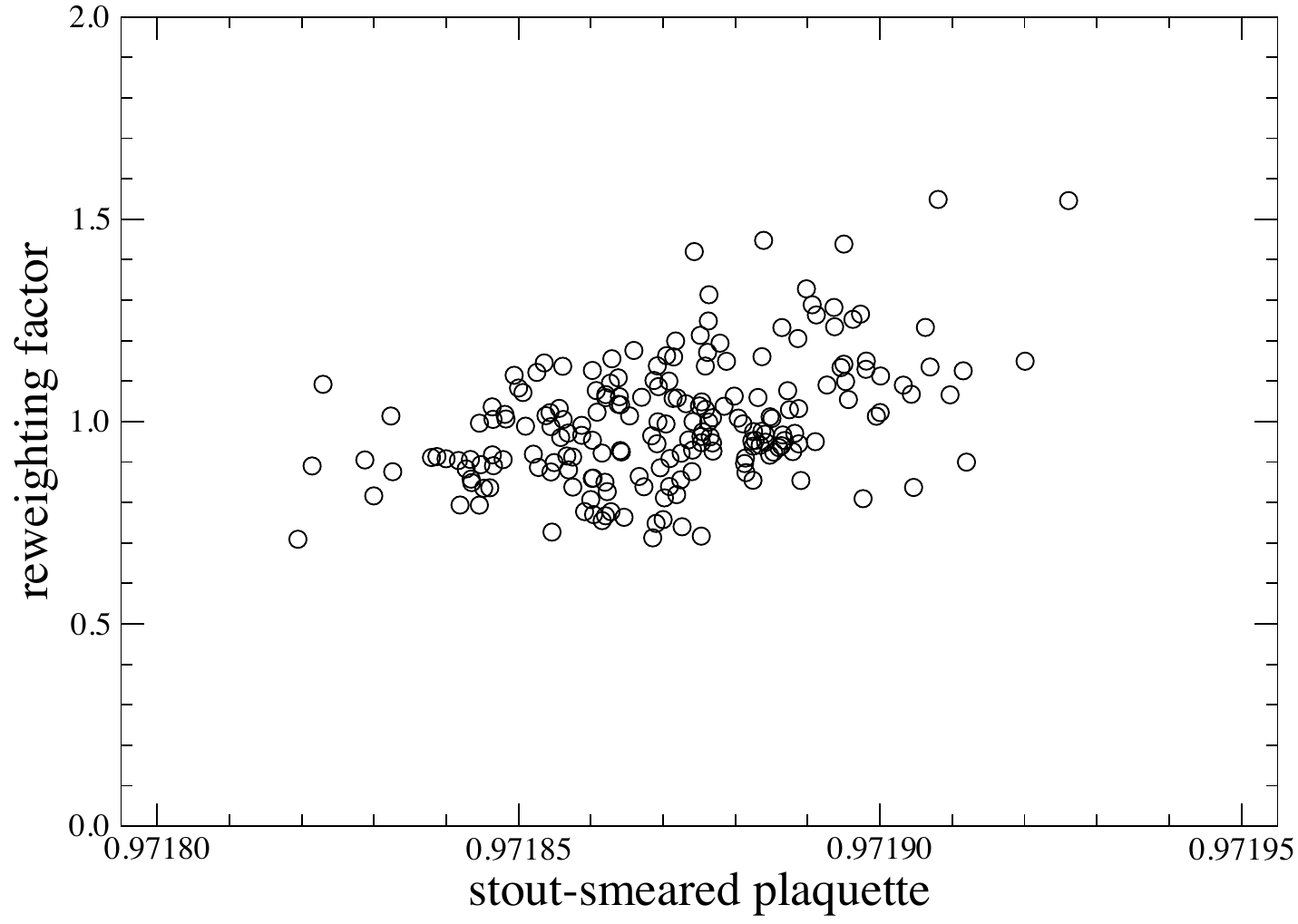}
\end{center}
\caption{Reweighting factor from 
$(\kappa_{\rm ud},\kappa_{\rm s})=(0.126117,0.124902)$ to 
$(\kappa_{\rm ud}^\target,\kappa_{\rm s}^\target)=(0.126119,0.124902)$
as a function of a stout-smeared plaquette value.}
\label{fig:rwfactor_p}
\end{figure}

\begin{figure}[h]
\begin{center}
\includegraphics[width=75mm,angle=0]{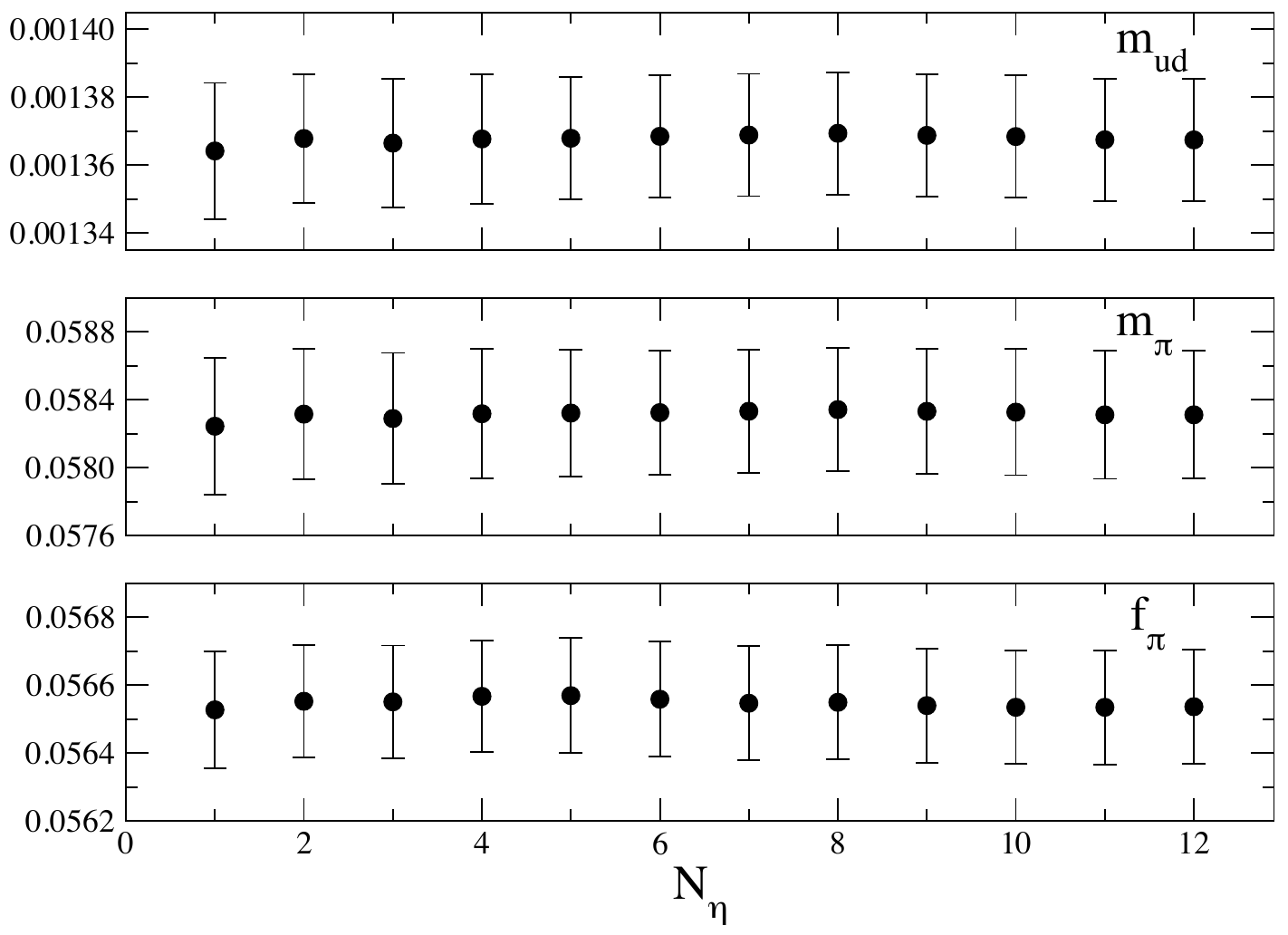}
\end{center}
\caption{Reweighted values for $m_{\rm ud}$ (top), $m_\pi$ (middle) and $f_\pi$ (bottom)  as a function of $N_\eta$.}
\label{fig:rw_noise}
\end{figure}

Let us turn to the analysis with the fixed AWI quark masses. 
In the previous section, we have found that the AWI quark masses on $64^4$ and $128^4$ lattices show deviation by 4.8{(1.6)}\%. We adjusted the AWI quark masses on the $64^4$ lattice to those on the $128^4$ lattice with the use of the reweighting technique explained in Sec.~\ref{subsec:reweighting}. The target hopping parameter is $(\kappa_{\rm ud}^\target,\kappa_{\rm s}^\target)=(0.126119,0.124902)$, which is obtained by a tiny shift of $(\Delta \kappa_{\rm ud}^\target,\Delta \kappa_{\rm s}^\target)=(+0.000002, \pm 0)$ from the simulation point. We choose $N_B^{\rm ud}=4$ for the number of the determinant breakup and introduce 12 sets of noise vectors for each determinant breakup. Figure~\ref{fig:rwfactor_c} shows the configuration dependence of the reweighting factor from $(\kappa_{\rm ud},\kappa_{\rm s})=(0.126117,0.124902)$ to $(\kappa_{\rm ud}^\target,\kappa_{\rm s}^\target)=(0.126119,0.124902)$, which is normalized by the configuration average. The fluctuations are less than 60\% around the average. In Fig.~\ref{fig:rwfactor_p}, we plot the reweighting factor as a function of the stout-smeared plaquette value on each configuration. We observe that the reweighting factor takes larger values as the plaquette value increases. This is an expected correlation, due to which the reweighted plaquette value at  $(\kappa_{\rm ud}^\target,\kappa_{\rm s}^\target)=(0.126119,0.124902)$ should be larger than the original one at $(\kappa_{\rm ud},\kappa_{\rm s})=(0.126117,0.124902)$. Figure~\ref{fig:rw_noise} tells us how many noise vectors are necessary to make the reweighted values converge. We observe that the values of  $m_{\rm ud}$, $m_\pi$, and $f_\pi$ with the error bars show little $N_\eta$ dependence. Similar behaviors are obtained for other physical quantities. So $N_\eta=12$ in our choice is sufficient.

In Table~\ref{tab:qmass}, we present the results for the reweighted AWI quark masses on $64^4$ lattice, which show good agreement with those on $128^4$ lattice both for the ud and s quarks. This assures us that the target hopping parameters are properly chosen.
In Fig.~\ref{fig:effmass}, the red triangle represents the reweighted effective pseudoscalar meson masses on $64^4$ lattice. We find that they are degenerate with those on $128^4$ lattice within the error bars for both the $\pi$ and $K$ mesons. The numerical values for the fit results in Table~\ref{tab:fitmass} give quantitative confirmation of the consistency. It is hard to detect the finite size effect on $m_\pi$ and $m_K$ between $64^4$ and $128^4$ lattices {under the current statistical precision}.  
On the other hand, the results for the PS meson decay constants are plotted in Fig.~\ref{fig:decayconst}. We find that the reweighted $f_{\pi}$ and $f_K$ show a clear finite size effect, whose magnitude is {0.66{(33)}\%} for $f_{\pi}$ and 0.26{(13)}\% for $f_K$. 

\subsection{Comparison with ChPT prediction}
\label{subsec:chpt}

\begin{figure}[th]
\vspace{3mm}
\begin{center}
\includegraphics[width=75mm,angle=0]{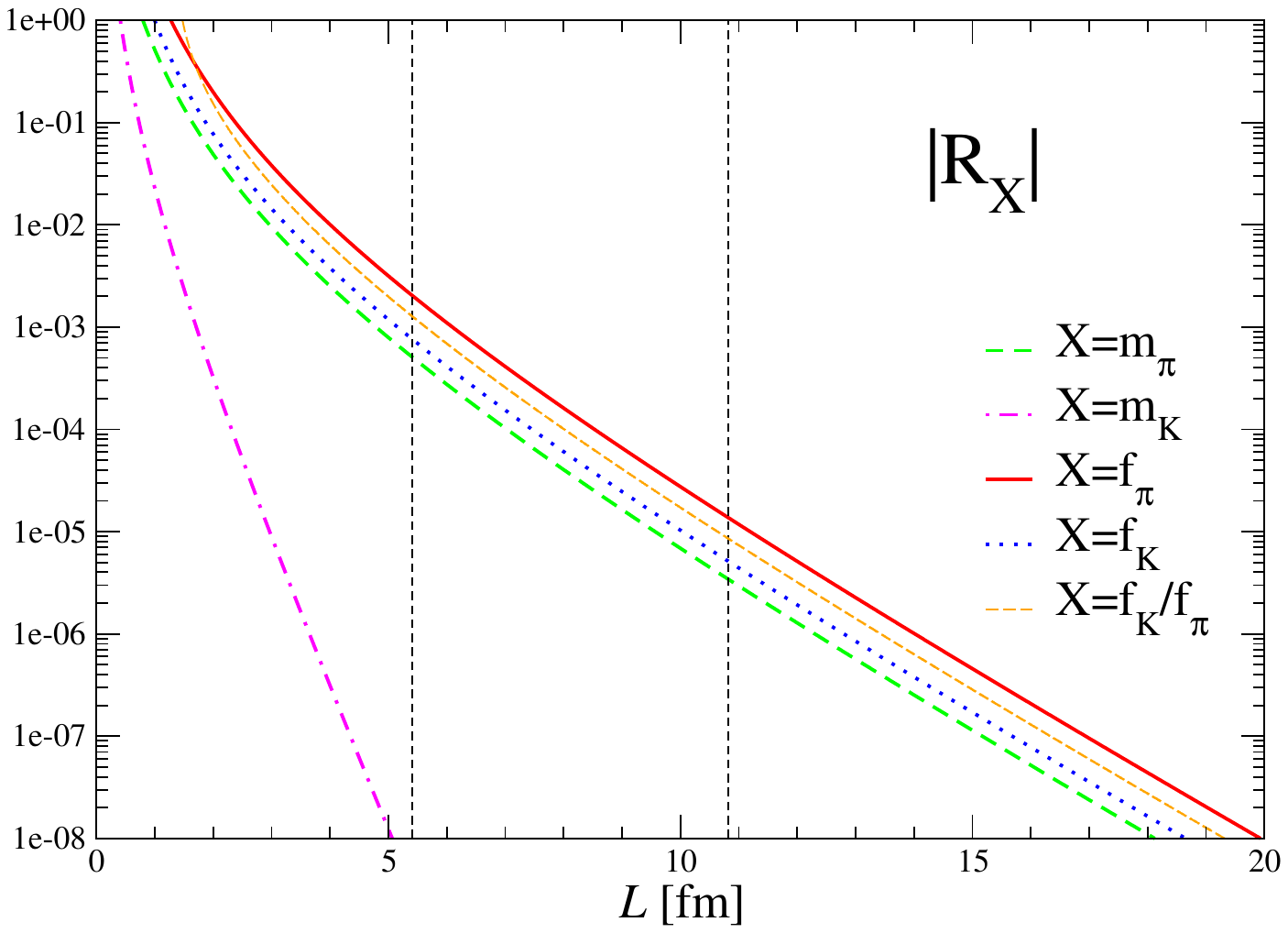}
\end{center}
\vspace{-.5cm}
\caption{$L$ dependence of $\vert R_X\vert $ $(R_{m_{\rm PS}}>0$ and $R_{f_{\rm PS}}<0)$ for $X=m_\pi,m_K,f_\pi,f_K,f_K/f_\pi $ at the physical point. Dotted vertical lines indicate $L=5.4$ fm and 10.8 fm.}
\label{fig:fse_chpt}
\end{figure}

In SU(3) ChPT, the full one-loop expressions for the finite size effects defined by $R_X=(X(L)-X(\infty))/X(\infty))$ for $X=m_\pi,m_K,f_\pi,f_K$ are given by~\cite{fse_chpt},
\ben
R_{m_\pi}&=&\frac{1}{4}\xi_\pi{\tilde g}_1(\lambda_\pi)-\frac{1}{12}\xi_\eta{\tilde g}_1(\lambda_\eta),\\
R_{m_K}&=&\frac{1}{6}\xi_\eta{\tilde g}_1(\lambda_\eta),\\
R_{f_\pi}&=&-\xi_\pi{\tilde g}_1(\lambda_\pi)-\frac{1}{2}\xi_K{\tilde g}_1(\lambda_K),\\
R_{f_K}&=&-\frac{3}{8}\xi_\pi{\tilde g}_1(\lambda_\pi)-\frac{3}{4}\xi_K{\tilde g}_1(\lambda_K)-\frac{3}{8}\xi_\eta{\tilde g}_1(\lambda_\eta)
\een
with
\ben
\xi_{\rm PS}&\equiv& \frac{2m_{\rm PS}^2}{(4\pi f_\pi)^2},\\
\lambda_{\rm PS}&\equiv& m_{\rm PS}L,\\
{\tilde g}_1(x)&\equiv&\sum_{n=1}^{\infty}\frac{4m(n)}{\sqrt{n} x}K_1(\sqrt{n} x),
\een
where $K_1$ is the modified Bessel function of the second kind and $m(n)$ denotes the multiplicity of the partition $n=n_x^2+n_y^2+n_z^2$. In Fig.~\ref{fig:fse_chpt}, we draw the $L$ dependence of $\vert R_X\vert $ $(X=m_\pi,m_K,f_\pi,f_K,f_K/f_\pi)$ at the physical point. Since we have a special interest in the values of $R_X $ at $L=5.4$ fm and $10.8$ fm, they are summarized in Table~\ref{tab:fse_chpt}. At $L=10.8$ fm, the expected finite size effect is at most 0.001\% found in $f_\pi$ channel, which is a completely negligible magnitude in the current numerical simulations. On the other hand, the lattice with $L=5.4$ fm can yield a detectable finite size effect, whose magnitude is expected to be 0.20\% for $f_\pi$, 0.08\% for $f_K$, and 0.13\% for $f_K/f_\pi$\footnote{The reason why it is hard to detect the finite size effect for $m_\pi$ is that its magnitude is expected to be smaller than those for $f_\pi$, $f_K$, and $f_K/f_\pi$ as found in Table IV, while the statistical error for $m_\pi$ is larger than those for $f_\pi$, $f_K$, and $f_K/f_\pi$.}. Actually, our simulation results support this feature semiquantitatively: {0.66{(33)}\%} finite size effect for $f_\pi$, 0.26{(13)}\% for $f_K$, and 0.40(32)\% for $f_K/f_\pi$, though they are roughly a factor of 3 larger than the ChPT predictions {with rather large error bars}. The discrepancy may be due to the higher-order effects in SU(3) ChPT, which are shown to enhance the magnitude of the finite size effects for $f_\pi$ and $f_K$  in Ref.~\cite{fse_chpt}.

{

It may be instructive to try another way to compare 
our results with the ChPT predictions.
The results for the pseudoscalar decay constants and the AWI quark masses
on $64^4$ lattice are interpolated onto the point where 
the reweighted pion mass reproduces the value predicted by the ChPT, 
which is lifted by 0.05\% from that on the $128^4$ lattice due to 
the finite size effects.  
Table~\ref{tab:fse_int} summarizes the interpolated results. 
They show little difference from the reweighted values at 
$(\kappa_{\rm ud}^\target,\kappa_{\rm s}^\target)=(0.126119,0.124902)$
in Tables~\ref{tab:fitmass}, \ref{tab:qmass}, and \ref{tab:fitdecayconst}.
This feature is also confirmed in Fig.~\ref{fig:decayconst} 
which compares the interpolated values and the reweighted ones 
for the pseudoscalar decay constants.
The difference between the interpolated results on $64^4$ lattice 
and  the simulated ones on $128^4$ lattice
is 0.63{(32)}\% for $f_\pi$,
0.25{(13)}\% for $f_K$, {and 0.39(32)\% for $f_K/f_\pi$}.

}

\begin{table}[h]
\caption{$R_X$ $(X=m_\pi,m_K,f_\pi,f_K,f_K/f_\pi)$ with $L=5.4$ fm and 10.8 fm at the physical point.}
\begin{center}
\begin{tabular}{crr}
\hline
$L$ & $64$ (reweighted) & $128$ \\\hline
$m_\pi L$ &   {3.73} & 7.48 \\
$R_{m_{\pi}}$ & $5.067\times 10^{-4}$ & $3.405\times 10^{-6}$ \\
$R_{m_{K}}$ & $3.317\times 10^{-9}$ & $1.912\times 10^{-16}$ \\
$R_{f_{\pi}}$ & $-2.027\times 10^{-3}$ & $-1.362\times 10^{-5}$ \\
$R_{f_{K}}$ & $-7.602\times 10^{-4}$ & $-5.108\times 10^{-6}$ \\
{$R_{f_{K}/f_\pi}$} & $ 1.269\times 10^{-3}$ & $8.513\times 10^{-6}$ \\\hline
\end{tabular}
\end{center}
\label{tab:fse_chpt}
\end{table}%
 
\begin{table*}[t]
\caption{Interpolated results on $64^4$ lattice at the point where the pion mass is lifted by 0.05\% from the value on $128^4$ lattice.}
\begin{center}
\begin{tabular}{cccccccc}
\hline
 $m_{\pi}$ & $m_{K}$ & $f_{\pi}$ & $f_{K}$ & $f_{K}/f_{\pi}$ & $m_{\rm ud}$ & $m_{\rm s}$ & $m_{\rm s}/m_{\rm ud}$\\\hline

 0.058460(373)  &  0.214611(114)  & 0.056555(167) & 0.067637(70) & 1.1960(35) & 0.001374(18) & 0.037999(10) & 27.66(36) \\\hline
\end{tabular}
\end{center}
\label{tab:fse_int}
\end{table*}%

\section{Conclusions and outlook}
\label{sec:conclusion}

We have investigated the finite size effect on the PS meson sector using (5.4 fm$)^4$ and (10.8 fm$)^4$ lattices in 2+1 flavor QCD at the physical point. The analysis at the fixed hopping parameters reveals 2.1{(8)}\%, 4.8{(1.6)}\%, and 0.36{(31)}\% finite size effect on $m_\pi$, $m_{\rm ud}$, and $f_\pi$, respectively. On the other hand, in the analysis at the fixed AWI quark masses with the aid of the reweighting technique, we detect {0.66{(33)}\%}, 0.26{(13)}\%, and {0.40(32)\%} finite size effect on $f_\pi$, $f_K$, {and $f_K/f_\pi$}, respectively. {These values with two-sigma error bars are consistent with the ChPT predictions of 0.20\%, 0.08\%, and 0.13\% for $f_\pi$, $f_K$ and $f_K/f_\pi$}.
We plan to make a similar finite size study for the baryon sector using the PACS10 configurations.

\begin{acknowledgments}
Numerical calculations are carried out on Oakforest-PACS through the HPCI System Research project (Project ID No.~hp170093, No.~hp180051) and the Interdisciplinary Computational Science Program in Center for Computational Sciences, University of Tsukuba. This work is supported in part by Grants-in-Aid for Scientific Research from the Ministry of Education, Culture, Sports, Science and Technology (MEXT) (Grants No.~16K13798, No.~16H06002, No.~18K03638). 
\end{acknowledgments}


\end{document}